\documentclass[aps,prl,twocolumn,showpacs]{revtex4}
\usepackage{epsfig}

\newcommand{\be}{\begin{equation}}
\newcommand{\ee}{\end{equation}}
\newcommand{\bea}{\begin{eqnarray}}
\newcommand{\eea}{\end{eqnarray}}

\begin{document}
\draft
\title{Effective Hamiltonian for ${\rm Ga}_{1-x}{\rm Mn}_x {\rm As}$ in the 
Dilute Limit}
\author{Gregory A. Fiete$^{1,2,3}$, Gergely Zar\'and$^{1,2,3}$ and
Kedar Damle$^{1,4}$
}
\address{
$^1$Department of Physics, Harvard University, Cambridge MA 02138 \\ 
$^2$Materials Science Division, Argonne National Laboratory, 9700
South Cass Avenue, Argonne IL, 60429\\ 
$^3$Research Institute of Physics,
Technical University Budapest, H-1521 Hungary\\
$^4$Department of
Physics and Astronomy, Rice University, Houston, TX 77005 }

\begin{abstract} 
We derive an effective Hamiltonian for ${\rm Ga}_{1-x}{\rm Mn}_x {\rm
As}$ in the dilute limit, where ${\rm Ga}_{1-x}{\rm Mn}_x {\rm As}$
can be described in terms of spin $F=3/2$ polarons hopping between the
{\rm Mn} sites and coupled to the local {\rm Mn} spins.  We determine
the parameters of our model from microscopic calculations. Our
approach treats the large Coulomb interaction in a non-perturbative
way, captures the effects of spin-orbit coupling and disorder, and is
appropriate for other p-doped magnetic semiconductors. Our model
applies to uncompensated {\rm Mn} concentrations up to $x \sim 0.03$.
\end{abstract}


\maketitle

Since their discovery\cite{ohno}, dilute III-V magnetic semiconductors
with high Curie temperatures have become the subject of very intense
research \cite{reviews}.  Because the magnetic ions (usually Mn)
responsible for the ferromagnetism are dissolved into the
semiconductor itself, these materials could provide a unique
opportunity to integrate ferromagnetic elements into larger,
non-magnetic, semiconducting devices.

In this Letter we focus on one of the most studied magnetic
semiconductors, ${\rm Ga}_{1-x}{\rm Mn}_x {\rm As}$, though most of
our calculations carry over to other p-doped III-V magnetic
semiconductors\cite{future}.  In ${\rm Ga}_{1-x}{\rm Mn}_x {\rm As}$
substitutional ${\rm Mn}^{2+}$ play a fundamental role: They provide
local spin $S=5/2$ moments, and they dope holes into the
lattice\cite{linnarsson}.  Since the ${\rm Mn}^{2+}$ ions are
negatively charged compared to ${\rm Ga}^{3+}$, in the very dilute
limit they bind these holes, forming an acceptor level with a binding
energy $E_b \approx 112 {\rm meV}$ \cite{linnarsson}.  As the ${\rm
Mn}$ concentration increases, these acceptor states start to overlap
and form an impurity band, which for even larger ${\rm Mn}$
concentrations merges with the valence band.  Though the actual
concentration at which the impurity band disappears is not known,
according to optical conductivity measurements\cite{singley}, this
impurity band seems to persist at least up to nominal ${\rm Mn}$
concentrations of about $x \approx 0.05$.  ARPES data\cite{asklund}
and the fact that even ``metallic'' samples feature a resistivity
upturn at low temperature\cite{vanesch} suggest that for smaller
concentrations (and maybe even for relatively large nominal
concentrations) one may be able to describe ${\rm Ga}_{1-x}{\rm Mn}_x
{\rm As}$ in terms of an impurity band \cite{bhatt2}.

In ${\rm Ga}_{1-x}{\rm Mn}_x {\rm As}$ the Coulomb potential created
by the ${\rm Mn}$ ions is by far the largest energy scale in the
problem \cite{vonOppen}, but spin-orbit coupling in the hole band is
also quite large compared to the exchange coupling between the holes
and the ${\rm Mn}$ spins\cite{linnarsson}.  Fortunately, the large
Coulomb potential of the $\rm Mn$ ion can be handled
non-perturbatively. We construct a many-body Hamiltonian in this limit
that captures spin-orbit effects, treats the large Coulomb interaction
non-perturbatively, and incorporates the exchange coupling between the
local moments and the holes.

The physics of the isolated ${\rm Mn}^{2+}$ + hole system is
well-understood \cite{linnarsson}: In the absence of the
$\rm{Mn}^{2+}$ core spin, the ground state of the bound hole at the
acceptor level is four-fold degenerate and well described in terms
of a $F=3/2$ spin.  For most purposes, we can restrict ourselves to
this fourfold degenerate $F=3/2$ acceptor level in the dilute limit.
As also evidenced by infrared spectroscopy \cite{linnarsson}, the
effect of the $S=5/2$ ${\rm Mn}$ core spin is well-described by a
simple exchange Hamiltonian \cite{reviews}:
\begin{equation}
H_{\rm exch} = G {\vec S} \cdot {\vec F}\;,
\end{equation}
with $G\approx5$ meV \cite{linnarsson}. 

The presence of other $\rm Mn$ sites has three important effects on
the $F=3/2$ acceptor state at any particular $\rm Mn$ site: {\em (i)}
The Coulomb potential of the neighboring ${\rm Mn}^{2+}$ ions will
induce a random {\em shift} $E$ of the fourfold degenerate
states. {\em (ii)} Because of the large spin-orbit coupling in GaAs,
the neighboring atoms will also generate an anisotropy $K$ and {\em
split} the fourfold {\em degeneracy} of the $F=3/2$ state into two
doubly degenerate states. {\em (iii)} Finally, the presence of the
neighboring ions will allow these $F=3/2$ spin objects to {\em hop}
between the ${\rm Mn}$ sites. However, this hopping $t$ will {\em not
conserve the spin} $F$ because of the large spin-orbit coupling. Thus,
in the dilute limit ${\rm Ga}_{1-x}{\rm Mn}_x {\rm As}$ should be
described by the following simple Hamiltonian:
\begin{eqnarray}
H  & = & \sum_{(i,j)} c^\dagger_{i,\mu}  t^{\mu \nu}_{ij}  c_{j,\nu} + \sum_i 
c^\dagger_{i,\mu} \;( K_i^{\mu\nu} + E_i \;\delta^{\mu\nu})\;
 c_{i,\nu} \nonumber \\
& + & G \sum_{i,\mu,\nu} {\vec S}_i \cdot (c_{i,\mu}^\dagger 
{\vec F}_{\mu\nu} \;c_{i,\nu})\;,
\label{eq:hamilt}
\end{eqnarray}
where $c^\dagger_{i,\nu}$ creates a hole at the acceptor level
$|F=3/2, F_z = \nu\rangle$ at position $i$. As is clear from the
arguments above, Eq.~(\ref{eq:hamilt}) is very general and appropriate
for describing other p-doped III-V and II-IV semiconductors as well
\cite{future}. Hole-hole interactions can be incorporated in
Eq.~(\ref{eq:hamilt}). In the metallic phase it is presumably a good
approximation to include only an on-site repulsion (discussed later)
that eliminates double occupancy of the acceptor levels. In the
localized phase, however, one may have to consider long-ranged
hole-hole interactions.
 
To estimate the various parameters in Eq.~(\ref{eq:hamilt}), we
studied the structure of the single impurity (Mn) and two-impurity
(${\rm Mn}_2$) bound hole states using the so-called spherical
approximation\cite{bald}.  The top of the valence band in ${\rm
Ga}_{1-x}{\rm Mn}_x {\rm As}$ can be described in terms of spin
$j=3/2$ holes \cite{kohn}, whose spin couples strongly to their
momenta. In the spherical approximation the motion of the holes in the
Coulomb potential of an ${\rm Mn}$ ion is described by\cite{bald},
\begin{equation}
H_0 = {\gamma \over 2  m }  \Bigl (p^2 - \mu \sum_{\alpha,\beta}
J_{\alpha\beta} p_{\alpha\beta} \Bigr ) - {e^2\over \epsilon\; r}
+ V_{cc}(r) \;,
\label{eq:h01}
\end{equation}
where $\gamma \approx 7.65$ is a mass renormalization parameter, $m$
is the free electron mass, $\mu \approx 0.77$ is the strength of the
spherical spin-orbit coupling in the $j=3/2$ band \cite{bald},
$\epsilon \approx 10$ is the dielectric constant of GaAs, and $V_{cc}$
is the so-called central cell correction \cite{guillaume}. The
spin-orbit term in Eq.~(\ref{eq:h01}) couples the momentum
tensor of the holes $p_{\alpha\beta} = p_\alpha p_\beta -
\delta_{\alpha\beta}\; p^2/3$ to their quadrupolar momentum,
$J_{\alpha\beta} = (j_\alpha j_\beta + j_\beta j_\alpha)/2 -
\delta_{\alpha\beta} \; j(j+1)/3$.

\begin{figure}[tb]
\begin{center}
\epsfxsize6cm
\epsfbox{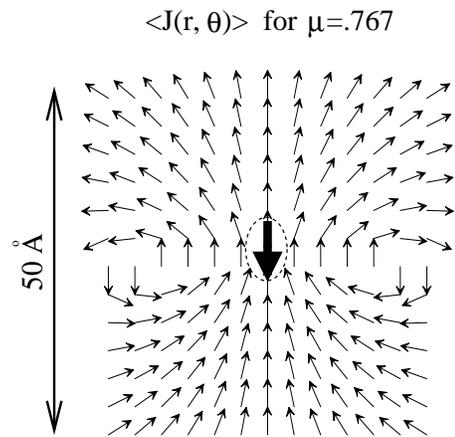}
\end{center}
\vskip0.1cm \caption{\label{fig:polar} Polarization of 
a bound hole in the state 
$|F=3/2, F_z = 3/2\rangle$ in $\rm Ga_{1-x}Mn_x As$ around a Mn ion (dark 
arrow pointing downwards represents the Mn S=5/2 spin).
Only the direction of the polarization is indicated. The magnitude falls 
off on a scale $\sim 10$~\AA.}
\end{figure}

The bound states of $H_0$ (without the central cell correction) have
been studied in the seminal paper\cite{bald}. Due to the spherical
symmetry, the total momentum, $\vec F \equiv \vec L + \vec j$, is a
conserved quantity, where $\vec L$ is the orbital angular momentum.
The ground state of $H_0$ is a four-fold degenerate $F=3/2$ multiplet
that contains a substantial d-wave contribution for {\rm GaAs} due to
the strong spin-orbit coupling.  In Fig.~\ref{fig:polar} we illustrate
the importance of this d-wave component by presenting the spatial
dependence of the hole polarization, $\vec j ({\bf r})$, for the state
$|F=3/2, F_z = 3/2\rangle$ which we calculated directly from the
Baldereschi-Lipari wave functions with the central cell correction.

\begin{figure}[tb]
\begin{center}
\epsfxsize7.5cm
\epsfbox{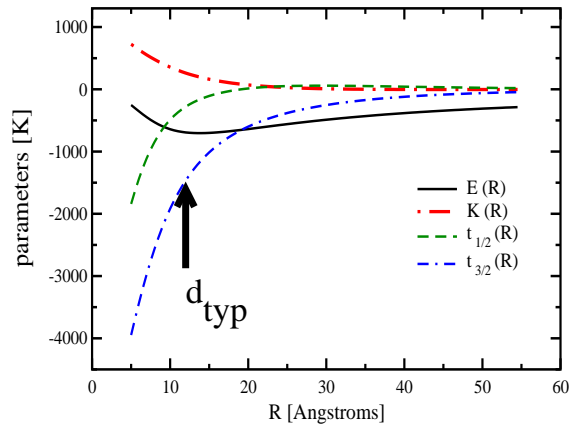}
\end{center}
\vskip0.1cm \caption{\label{fig:couplings} Parameters of the
two-impurity Hamiltonian Eq.~(\ref{eq:two_impurity}) obtained from the
variational study of two Mn ions.  The arrow indicate the typical
Mn-Mn distance, $d_{\rm typ}$, for $x=0.01$ Mn concentration.  }
\end{figure}

Having computed the single ${\rm Mn}$ hole states, we carried out a
variational calculation to construct the molecular orbitals for a pair
of ${\rm Mn}$ ions \cite{future,bhatt3}.  Since the exchange
interaction with the Mn core spins is much smaller than the binding
energy of the holes, we neglected the effect of $G$ on the their wave
functions in these calculations.  For the case where both the Mn-Mn
bond and the quantization axis of $F$ are parallel to the $z$-axis,
$F_z$ is conserved and the spectrum of the lowest lying states of the
molecule can be fully characterized by:
\begin{eqnarray}
H^{\rm eff}_{{\rm Mn}-{\rm Mn}}& = &\sum_\nu t_\nu(R) 
\left( c^\dagger_{1,\nu}c_{2,\nu} + {\rm h.c.} \right ) 
\label{eq:two_impurity}\\
&+& \sum_{\scriptstyle i=1,2 \atop \scriptstyle\nu}  \left( K(R) \; \left(\nu^2
-{5\over 4}\right ) + E(R) \right)  \; c^\dagger_{i,\nu} c_{i,\nu}
\;.
\nonumber \\
\end{eqnarray}
By time reversal symmetry, the hopping parameters satisfy $t_{3/2} =
t_{-3/2}$ and $t_{1/2} = t_{-1/2}$.  All parameters depend only on the
distance $R$ between the two Mn sites (see Fig.~\ref{fig:couplings}).
The most obvious effect of the spin-orbit coupling is that the
hoppings $t_{3/2}$ and $t_{1/2}$ substantially differ from each-other;
holes that have their spin aligned with the Mn-Mn bond are more
mobile.  As indicated by the arrow, at the typical Mn-Mn distance for
$x=0.01$, $K$ and $t_{1/2}$ can be entirely neglected compared to 
$E$ and $t_{3/2}$. Therefore, in many cases it is enough to keep only the
latter two terms in the effective Hamiltonian.

Having determined the effective Hamiltonian for  a pair of Mn ions,
we can use it to estimate the parameters in Eq.~(\ref{eq:hamilt}). 
Rotating the $z$-axis along the bond direction ${\vec n}_{ij}$ connecting 
sites $i$ and $j$, we obtain ${\bf t}_{ij}  =   
{\bf \cal D}({\vec n}_{ij}) {\bf \hat t}(R_{ij}) {\bf\cal D}^\dagger 
({\vec n}_{ij}) $, 
${\bf K}_i  =  {1\over 2} \sum_{j\ne i}  K(R_{ij})
\bigl(({\vec n}_{ij}\cdot  {\vec F})^2 - {5\over 4}\bigr) $, and
$E_i = {1\over 2} \sum_{j\ne i} E(R_{ij})$. Here 
${\bf \cal D}({\vec n}_{ij})$ is a spin 3/2 rotation matrix,  and 
${\bf \hat t}(R)$
denotes the diagonal matrix $diag \left(t_{3/2} (R),\; t_{1/2} (R),\; 
t_{1/2} (R),\; t_{3/2} (R)\right)$, and $R_{ij}$ denotes the distance 
between sites $i$ and $j$. 

So far we have neglected the interaction between holes.  In the
localized phase, however, this interaction may play an important role.
In general, the Coulomb interaction between holes on different Mn
sites has a very complicated form \cite{future}, though for large
separations it simplifies considerably. Fortunately, the dominant
interaction is the {\em on site} hole-hole interaction. Within the
spherical approximation this interaction can be expressed as:
\begin{equation} 
H_{\rm int} = {U_N\over 2} \sum_i :{{\hat N}_i}^2:
+  {U_F\over 2} \sum_i :{\hat {\vec F}}_i^2:\;,
\label{eq:int}
\end{equation}
where ${\hat N}_i= \sum_\nu c^\dagger_{i,\nu}c_{i,\nu}$, ${\hat {\vec
F}}_i = \sum_{\mu,\nu} c^\dagger_{i,\mu}{\vec F}_{\mu\nu} c_{i,\nu}$,
and $:...:$ denotes normal ordering. We estimated $U_N$ and $U_F$ in
Eq.~(\ref{eq:int}) by evaluating exchange integrals: $U_N
=2600 $K and $U_F=-51$ K.

Eqs.~(\ref{eq:hamilt}) and (\ref{eq:int}), together with the
microscopic parameters of Fig.~\ref{fig:couplings}, constitute our
central results. They provide a well controlled theoretical framework
that captures the most important aspects of dilute magnetic
semiconductors such as the localization phase transition, random
anisotropy, disorder effects, and frustrated
ferromagnetism. Postponing much of our detailed analysis to a longer
publication\cite{future}, here we only demonstrate the power of this
model on a few examples.  

\begin{figure}[tb]
\begin{center}
\epsfxsize7.5cm
\epsfbox{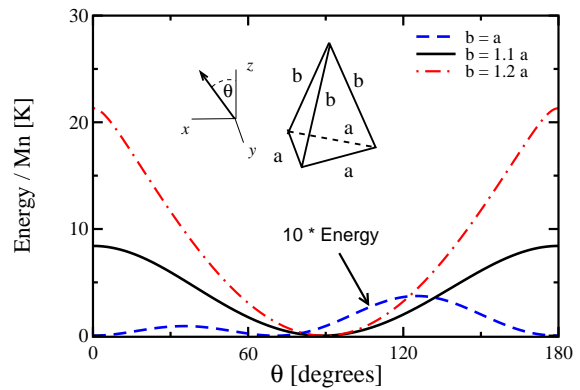}
\end{center}
\vskip0.1cm \caption{\label{fig:aniso} Anisotropy induced by the distortion 
of a regular Mn tetrahedron in the presence of a single hole. 
The Mn-Mn distances are $a=15\AA$ and $b=15 \AA$, $b=16.5 \AA$,
and $b=18 \AA$, respectively. The distortion generated 
anisotropy can be almost two orders of magnitude larger 
than the undistorted anisotropy, which is of order 1K/Mn.
} 
\end{figure}

To obtain a better understanding of the model we first computed the
ground state of four Mn atoms at a separation of $15\AA$ due to the
interaction mediated by a single hole on the cluster.  We treated the
Mn spins classically and used the simple mean field approximation of
Ref.~\cite{bhatt2}.  We considered only configurations where the Mn
ions were positioned on a slightly distorted tetrahedron with three
edges of length $a=15 \AA$ and three edges of length $b$ (see
Fig.~\ref{fig:aniso}).  In all cases, in the ground state, the Mn
spins are relatively collinear apart from a slight tilt of $5-10^{o}$.
However, the spatial position of the Mn ions generates a strong
anisotropy. Thus, the energy depends strongly on the directional
orientation of the net spin relative to the underlying lattice.  To
demonstrate this we calculated the ground state energy as a function
of the Mn spin direction (assuming full alignment).  For a perfectly
regular tetrahedron this anisotropy is rather small, less than $0.5
{\rm K}/ {\rm Mn}$. However, the anisotropy increases with the ratio
$b/a$, and for $b/a = 1.2$ it can be as large as $20 {\rm K} / {\rm
Mn}$. In other words, random positions of the Mn ions induce a {\em
random anisotropy} term that, depending on the disorder, is much
larger than the bulk anisotropy, which is of the order of $1 {\rm K}/
{\rm Mn}$.  Thus disorder and spin-orbit coupling together can induce
a large random anisotropy energy comparable to $T_C$.  These findings
are in good agreement with earlier results obtained in the metallic
limit \cite{gergely}.

\begin{figure}[tb]
\begin{center}
\epsfxsize7.3cm
\epsfbox{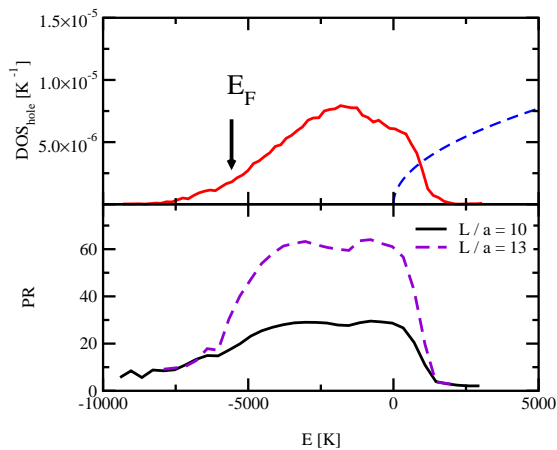}
\end{center}
\vskip0.1cm \caption{\label{fig:lattice} Top: Computed average hole
density of states for 10 $L= 10 a$ samples with $x=0.01$ and $f=0.3$.
We also show the density of states of the valence band (dashed
line). The Fermi energy is $\approx -6500$ K.  Bottom: The
participation ratio for $L = 10 a_{\rm lat}$ and $L=13a_{\rm lat}$.
States in the impurity band tails are localized while states in the
middle are delocalized; states in the side tail above zero energy likely mix
with the valence band states and are delocalized in reality.  }
\end{figure}

Finally, we discuss some of the results obtained for a $\rm
Ga_{1-x}Mn_x As$ of linear sizes $L=10 a_{\rm lat}$ and $L=13 a_{\rm
lat}$ (with $a_{\rm lat}=5.65 \AA$, the size of the conventional unit
cell) and active Mn concentration $x_{\rm active}=0.01$, using the
above-described mean field techniques at zero temperature.  In the
calculations presented below, we have not included the effects of
Eq.~(\ref{eq:int}). (This is partially justified post-facto by
Fig.~\ref{fig:lattice} which shows the states at the Fermi energy are
delocalized. If these states were localized one would expect the
contributions of Eq.~(\ref{eq:int}) to be more important.)  We
emphasize that $x_{\rm active}$ can be substantially less then the
nominal Mn concentration, $x$, which also includes inactive Mn sites
\cite{yu}, and therefore these calculations may be relevant even for
systems with larger nominal Mn concentration. The concentration of
holes is also reduced compared to $x$ due to strong compensation
effects; we assumed that the number of holes is reduced by a factor of
$f=0.3$ relative to the number of Mn.

To take into account correlations\cite{vonOppen} induced between Mn
ions during the experimental growth process, we introduced a screened
Coulomb repulsion between the Mn ions and let them relax using T=0
Monte Carlo simulations. For long times the Mn ions form a regular bcc
lattice with some point defects. The data we present here are for
intermediate times, where there is still appreciable disorder in the
system.

Once we fixed the Mn positions in a given instance, we solved the mean
field equations derived from (\ref{eq:hamilt}) self consistently
\cite{bhatt2}.  We used periodic boundary conditions and implemented a
short distance cutoff in the hopping parameters of
Eq.~(\ref{eq:hamilt}) which corresponds to about 8 neighbors for each
Mn. The use of this cut-off is justified by the observation that our
molecular orbital calculations are only appropriate for ``nearest
neighbor'' ion pairs, and in reality, holes cannot hop directly over
the first ``shell'' of ions. We started from a configuration with
fully alligned classical Mn spins, $\vec \Omega_{i} \equiv \vec S /
S$, and then let the system relax to the nearest metastable state.
Similar to the metallic case \cite{gergely}, we find a ferromagnetic
state with a largely reduced magnetization, $|\langle \vec
\Omega_i\rangle|\approx 0.4$ for $L= 10 a_{\rm lat}$.  We find that
this reduction is largely due to spin-orbit coupling, and that the
cosine of the angle $\theta$ between the spins and the ground state
magnetization has a broad distribution similar to the metallic case
\cite{gergely}.

The density of states is shown in Fig.~\ref{fig:lattice}.  The
halfwidth of the impurity band is about $0.2$ eV at this density,
which slightly overlaps with the valence band density of states.
However, comparison with the valence hole density of states suggests
that at this concentration a well-formed impurity band may still be
present, and it might persist to higher concentrations.  Indeed, this
scenario seems to be supported by many
experiments\cite{singley,asklund}.

The impurity band has a tail of localized states that reaches 
inside the band gap. These states can be identified for
various system sizes. (See Fig.~\ref{fig:lattice} wherein the
participation ratio, ${\rm PR}= [ \sum_i (\sum_\alpha
\vert\psi_{i\alpha}\vert^2)^2]^{-1}$, grows with system size for
delocalized states while the PR remains O(1) in the thermodynamic
limit for localized states).

This tail gradually disappears when we introduce correlations between
the Mn ions which tend to form regular structures \cite{future}.  In
agreement with ARPES data\cite{asklund}, we find that the
chemical potential lies deep ($\sim 0.5\, {\rm eV}$) inside the
gap. From the PR data, it appears that the chemical potential is
in the vicinity of the mobility edge, a regime where our model is
probably more reliable.  This raises the interesting possibility that
the localization phase transition in ${\rm Ga}_{1-x}{\rm Mn}_x {\rm
As}$ could happen inside the impurity band and that the ferromagnetic
phase for smaller Mn concentrations is governed by localized hole
states \cite{vonOppen,yang}.

Though our calculations are based on microscopic model calculations,
they are only approximate, and more realistic {\it ab initio}
calculations would be needed to give a quantitative answer concerning
the role of the impurity band. Also, though the spherical
approximation we used is able to reproduce rather well the spectrum of
a single acceptor, it might overestimate the effect of spin-orbit
coupling, and also the width of the impurity band.

In summary, based on microscopic calculations we constructed a
many-body Hamiltonian that is appropriate for describing ${\rm
Ga}_{1-x}{\rm Mn}_x {\rm As}$ in the dilute limit. We find that the
hopping of the carriers is strongly correlated with their spin.  This
spin-dependent hopping is crucial for capturing spin-orbit coupling
induced random anisotropy terms, the lifetime of the magnon
excitations, and for capturing the universality class of the
localization phase transition. Our calculations suggest the presence
of an impurity band for $x_{\rm active} = 0.01$ Mn concentration.

We are grateful to B. Jank\'o, M. Berciu, R. Bhatt, A. H. MacDonald,
P. Schiffer, and especially J.K. Furdyna, X. Liu, and E. Sasaki for
stimulating discussions. This research has been supported by the
U.S. DOE, Office of Science, NSF Grants No. DMR-9985978 and
DMR97-14725, and Hungarian Grants No. OTKA F030041, T038162, and
N31769.


\vspace{-0.8cm}


\begin{references}
\vspace{-.7cm}
\bibitem{ohno}H. Ohno, Science {\bf 281} 951 (1998).
\bibitem{reviews} J. K\"onig {\em et al.}  in
Electronic Structure and Magnetism of Complex Materials, Eds.
D.J. Singh and D.A. Papaconstantopoulos (Springer Verlag 2002);
R. N. Bhatt {\em et al.}, J.  Superconductivity INM {\bf 15}, 71
(2002); T. Dietl, cond-mat/0201282.
\bibitem{future} G. Fiete, G. Zar\'and and K. Damle (unpublished).
\bibitem{linnarsson} M. Linnarsson {\em et al.}, Phys. Rev. B {\bf
55}, 6938 (1997);
\bibitem{singley} E. J. Singley {\em et al.}, Phys. Rev. Lett. {\bf
89} 097203 (2002).
\bibitem{asklund} H. \AA sklund {\em et al.}, Phys. Rev. B {\bf 66}
115319 (2002).
\bibitem{vanesch} A. Van Esch {\em et al.}, Phys. Rev. B {\bf 56}
13103 (1997).
\bibitem{bhatt2} M. P. Kennett, M. Berciu and R. N. Bhatt,
Phys. Rev. B {\bf 66} 045207 (2002).
\bibitem{vonOppen} C. Timm, F. Sch\"afer and F. von Oppen,
Phys. Rev. Lett {\bf89} 137201 (2002).
\bibitem{bald}
A. Baldereschi, and N.O. Lipari, Phys. Rev. B {\bf 8}, 2697
(1973).
\bibitem{kohn}W. Kohn and J.M. Luttinger,  Phys. Rev. 98, 915 (1955).
\bibitem{guillaume} A. K. Bhattacharjee and C. Benoit \'a la
Guillaume, Solid State Comm. {\bf 113}, 17 (2000).
\bibitem{yang} S.-R. Yang and A. H. MacDonald, Phys. Rev. B {\bf 67}, 155202
(2003).
\bibitem{bhatt3} A.C. Durst, R.N. Bhatt and P.A. Wolff, Phys. Rev. B {\bf 65},
235205 (2002).
\bibitem{gergely} G. Zar\'and and B. Jank\'o, 
Phys. Rev. Lett. {\bf 89}, 047201  (2002). 
\bibitem{yu} K. M. Yu, W. Walukiewicz {\em et al.}, Appl. Phys. Lett. {\bf 81},
844 (2002), K. M. Yu {\em et al.}, Phys. Rev. B {\bf 65}, 201303 (2002).
\end{references}
\end{document}